\documentstyle[12pt,aasms4]{article}

\newdimen\sa  \newdimen\sb
\def\pdeg{\ifmmode $\setbox0=\hbox{$^{\circ}$}\rlap{\hskip.11\wd0 .}$^{\circ}
          \else \setbox0=\hbox{$^{\circ}$}\rlap{\hskip.11\wd0 .}$^{\circ}$\fi}
\def\gtorder{\mathrel{\raise.3ex\hbox{$>$}\mkern-14mu
             \lower0.6ex\hbox{$\sim$}}}
\def\ltorder{\mathrel{\raise.3ex\hbox{$<$}\mkern-14mu
             \lower0.6ex\hbox{$\sim$}}}
\newcommand{\Ho}{\mbox{$H_0$}}
\newcommand{\kms}{\mbox{ km~s$^{-1}$}}
\newcommand{\kmm}{\mbox{ km~s$^{-1}$ Mpc$^{-1}$}}

\newcommand{\lsolar}{\mbox{$L_\odot$}}

\slugcomment{Submitted to the Astrophysical Journal}

\lefthead{Impey et al.}
\righthead{An Einstein Ring in the Gravitational Lens System PG 1115$+$080}

\begin{document}

\title{An Infrared Einstein Ring in \\
  the Gravitational Lens PG~1115+080\footnote{Based on Observations made 
  with the NASA/ESA Hubble Space Telescope, obtained at the Space Telescope 
  Science Institute, which is operated by AURA, Inc., under NASA contract 
  NAS 5-26555. } }

\vskip 2truecm

\author{C. D. Impey} 
\affil{Steward Observatory, University of Arizona, Tucson, AZ 85721}
\affil{email: cimpey@as.arizona.edu}

\author{E. E. Falco, C. S. Kochanek, J. Leh\'ar, and B. A. McLeod}
\affil{Harvard-Smithsonian Center for Astrophysics, 60 Garden St., Cambridge,
	MA 02138}
\affil{email: falco@cfa.harvard.edu, ckochanek@cfa.harvard.edu, 
	jlehar@cfa.harvard.edu, bmcleod@cfa.harvard.edu}

\author{H.-W. Rix and C. Y. Peng}
\affil{Steward Observatory, University of Arizona, Tucson, AZ 85721}
\affil{email: rix@as.arizona.edu, cyp@as.arizona.edu}

\and 

\author{C. R. Keeton}
\affil{Harvard-Smithsonian Center for Astrophysics, 60 Garden St., Cambridge,
	MA 02138}
\affil{email: ckeeton@cfa.harvard.edu}


\begin{abstract}

Hubble Space Telescope observations of the gravitational lens PG~1115+080 
in the infrared show the known $z_l=0.310$ lens galaxy and reveal the
$z_s=1.722$ quasar host galaxy. The main lens galaxy G is a nearly circular
(ellipticity $\epsilon < 0.07$) elliptical galaxy with a de Vaucouleurs 
profile and an effective radius of $R_e = 0\farcs59\pm0\farcs06$ ($1.7 
\pm 0.2 h^{-1}$ kpc for $\Omega_0=1$ and $h = H_0/100$ \kmm). G is part 
of a group of galaxies that is a required component of all successful
lens models. The new quasar and lens positions (3 milliarcsecond errors) 
yield constraints for these models that are statistically degenerate,
but several conclusions are firmly established. (1) The principal lens
galaxy is an elliptical galaxy with normal structural properties, lying 
close to the fundamental plane for its redshift. (2) The potential of
the main lens galaxy is nearly round, even when not constrained by 
the small ellipticity of the light of this galaxy. (3) All models
involving two mass distributions place the group component near the
luminosity-weighted centroid of the brightest nearby group members.
(4) All models predict a time delay ratio $r_{ABC}\simeq 1.3$. (5) Our
lens models predict $H_0=44\pm4$ \kmm\ if the lens galaxy contains
dark matter and has a flat rotation curve, and $H_0=65\pm5$ \kmm\ if
it has a constant mass-to-light ratio. (6) Any dark halo of the main
lens galaxy must be truncated near $1\farcs5$ ($4 h^{-1}$ kpc) before
the inferred \Ho\ rises above $\sim 60$ \kmm. (7) The quasar host
galaxy is lensed into an Einstein ring connecting the four quasar
images, whose shape is reproduced by the models. Improved NICMOS
imaging of the ring could be used to break the degeneracy of the lens
models.

\end{abstract}


\section{INTRODUCTION}

Gravitational lens time delays offer a means of determining the Hubble
constant that is purely geometrical and hence completely avoids the
complications of the local distance
scale (Refsdal 1964). The time delay for Q~0957+561 is now well-measured
(Schild \& Thomson 1997; Kundi\'c et al. 1997; Haarsma et al. 1997), but
significant systematic uncertainties remain due to the degeneracy between 
the mass of the primary lens galaxy and its host cluster (e.g. Grogin \& 
Narayan 1996; Bernstein et al. 1997; Romanowsky \& Kochanek 1998). No 
single lens is likely to be completely free of systematic uncertainties, 
so a reliable estimate of $H_0$ should rely on an ensemble of lenses. 
There are now three more systems with time delay estimates: PG~1115+080
(Schechter et al. 1997), B~1608+656 (Fassnacht et al. 1996), and B~0218+357
(Corbett et al. 1996), which need detailed exploration of their lens models 
to examine the systematic uncertainties.  

PG~1115+080 was the second gravitationally lensed quasar to be
discovered (Weymann et al. 1980). The source is an optically
selected, radio-quiet quasar at redshift $z_s=1.722$. Hege et al. 
(1981) first resolved the four quasar images (a close pair A1/A2, B
and C), confirming the early model of Young et al. (1981) that the
lens was a five-image system, one image being hidden in the core of
the lens galaxy. Henry \& Heasly (1986) detected the lens galaxy,
followed by gradual improvements in the astrometry by Kristian et
al. (1993; hereafter K93), and Courbin et al. (1997). The redshift of
the lens galaxy was determined by Angonin-Willaime, Hammer \& Rigaut
(1993) and confirmed by Kundi\'c et al. (1997) and Tonry (1998) to be
$z_l=0.310$. Tonry also determined the central velocity dispersion of 
the lens galaxy: $\sigma = 281\pm25$ km s$^{-1}$. The spatial resolution 
of published data has always been insufficient to perform any surface
photometry on the lens galaxy. Young et al. (1981) noted that
the lens seemed to be part of a small group centered to the southwest
of the lens, with a velocity dispersion of approximately $270\pm70$ km
s$^{-1}$ based on only four galaxy redshifts (Kundi\'c et al. 1997). The 
group is an essential component of any model that successfully fits 
the lens constraints (Keeton, Kochanek \& Seljak 1997; Schechter et al.
1997). Finally, Schechter et al. (1997) successfully determined two 
time delays between the images, which were reanalyzed by Barkana 
(1997) to give $\Delta\tau_{BC}=25.0^{+1.5}_{-1.7}$ days and the 
time delay ratio $r_{ABC}=\Delta\tau_{AC}/\Delta\tau_{BA}=1.13\pm0.18$.  
These results were analyzed by Keeton \& Kochanek (1997) and Courbin 
et al. (1997) to deduce $H_0=53_{-7}^{+15}$ km s$^{-1}$ Mpc$^{-1}$, 
with comparable contributions to the uncertainties from the time delay
measurement and the models. The extreme variations are given in 
non-parametric form by Saha \& Williams (1997), although some of 
these models may not be physical. 

We present new near-infrared observations of the PG~1115+080 system
obtained with the Hubble Space Telescope (HST) NICMOS camera. These 
are the first results of the CfA-Arizona Space Telescope Lens Survey 
(CASTLES).\footnote{A summary of gravitational lens data and model
results, including CASTLES data, is available at the URL
http://cfa-www.harvard.edu/castles.} After 
summarizing the observations in \S2, we present improved astrometry 
in \S2.1, the first surface photometry of the lens galaxy in \S2.2, 
a discussion of lens models and the Hubble constant in \S2.3, photometry 
of the nearby group in \S2.4, and comments on the quasar host galaxy in 
\S2.5. In \S3 we comment on the strengths and limitations of this system 
as a cosmological tool.

\section{RESULTS}

The HST observations were made on 17 November 1997 using NICMOS Camera
2 and the $H$ (F160W) filter. Four images were taken in a spiral dither
pattern for a total integration time of 2560 seconds. The field of
view is $\sim 19\arcsec\times 19\arcsec$. We reduced the data with a 
modified pipeline (McLeod 1998). Figure 1a shows the sum of the dithered,
flat-fielded F160W images, and Figure 1b shows the image after the quasar 
point sources were fitted and subtracted. Figures 1c and 1d show model 
results that are discussed below. 

We also reanalyzed the WFPC1 images of K93 from 3 March 1991, which 
consisted of a 60 second $V$ (F555W) exposure and two 400 second $I$ 
(F785LP) exposures, and also a deeper set of unpublished WFPC1 images 
by Westphal (obtained from the STScI archives; hereafter W93) taken on 
18 February 1993 and consisting of a total exposure of 4400 seconds in the 
F555W filter and 7200 seconds in the F785LP filter. The field of view of 
the images is $\sim 35\arcsec\times35\arcsec$. Figure 2 shows the combined
W93 F785LP image scaled to reveal the galaxies making up the 
surrounding group at the expense of showing the lens geometry.

\subsection{Astrometry}

We obtained relative astrometry for the quasar images and the lens galaxy 
using both detailed model fitting (with two independent implementations)
and centroiding. The various estimates for the component positions 
relative to quasar image C in the NICMOS image are consistent to an 
rms internal measurement error of 2 milliarcseconds (mas), consistent 
with the variations found by bootstrap error estimation using random 
combinations of the four NICMOS exposures. The uncertainties are dominated 
by the effects of binning the images into the approximately 76 mas 
(two-times oversampled at 38 mas) pixels. We also fitted both the K93 
images and the unpublished W93 images. Our fits to the K93 data differ 
significantly from those in K93, but they are consistent with our 
fits to the W93 images. The astrometry is listed in Table 1.

The most general global coordinate transformation between the systems 
is of the form

\begin{equation}
  T = (1-\kappa) \left( \begin{array}{cc}
                            \cos\theta & -\sin\theta \\ 
                            \sin\theta &  \cos\theta 
                        \end{array}
                  \right)
        + \gamma \left( \begin{array}{cc}
                            \cos2\theta_\gamma &  \sin2\theta_\gamma \\
                            \sin2\theta_\gamma & -\cos2\theta_\gamma 
                        \end{array}
                  \right)
\end{equation}
\noindent
where $1-\kappa$ represents a scale change, $\theta$ 
represents a global rotation, 
and $\gamma$ and $\theta_\gamma$ 
represent a relative shearing of the coordinates. 
We have written the transformation in a form that closely corresponds to 
terms that appear in the gravitational lens constraint equations
(e.g., Schneider, Ehlers \& Falco 1992). The 
scale change term is equivalent to adding a convergence $\kappa$ between 
the two solutions; it will produce absolute fractional changes in the 
model parameters and inferred Hubble constant equal to $\kappa$.  
Global rotations will have no effect on the solutions because we have no 
constraints that depend on the absolute orientation. Coordinate shears 
are roughly equivalent to the effects of tidal gravity and ellipticity 
but alter the solutions in a non-trivial way. Fits of the transformation 
matrix demonstrated that the differences between the astrometric 
solutions in Table 1 were 
overwhelmingly dominated by a scale change and a rotation rather than 
a shear or random errors.  

For the astrometric results 
in Table 1, we adopted pixel scales of $0\farcs076030$ 
in the X direction and $0\farcs075344$ in the Y direction for the NIC2 
camera (Cox et al. 1997; from the measurement closest in time to our
observations). For the WFPC1 data of Westphal, we adopted the pixel scale 
of $0\farcs04404$ from the image headers. Note that this value 
differs both from 
the Gould \& Yanny (1994) scale of $0\farcs04374$ and the older calibration 
of $0\farcs04389$ used by K93. For these latter two pixel scales, the 
NICMOS and WFPC1 astrometry differed at the level of 20 mas, which was 
far larger than any plausible source of errors (Gilmozzi et al. 1995). 

With the adopted WFPC1 pixel scale of $0\farcs04404$, the rms difference 
between the W93 and NICMOS astrometry is 6 mas. If we allow a small 
rotation of $\theta \approx 0\pdeg1$ 
between the images due to the uncertainties in the 
absolute orientation of the images ($0\pdeg03$ to $0\pdeg05$ rms, see 
Lupie et al. 1997), the rms differences are only 5 mas. If in addition 
we allow a small scale difference of $\kappa \simeq 0.0032$, the rms 
differences are only 1 mas. We finally adopted for our models an 
uncertainty of 3 mas per coordinate, or slightly over 4 mas per relative
coordinate, corresponding to roughly twice the random errors and four 
times the systematic errors. We also checked the NICMOS astrometry by 
comparing our position estimates for MG~0414+0534 to VLBI positions 
(using component b of Trotter 1998), and by comparing our position 
estimates for H~1413+117 to the ground based astrometry of Schechter 
(private communication), and in both cases found consistent results 
at the $\leq 3$ mas level.  

\subsection{Photometry}

Table 2 summarizes the photometry for all observations that could
reliably detect the lens galaxy.  The NICMOS photometry was obtained
by fitting a model to the image using synthetic 
point-spread functions (PSFs) we generated with Tiny Tim 
(v4.4; Krist \& Hook 1997). We adopt a F160W zeropoint of 1087 Jy 
for zero magnitudes at an effective wavelength of 1.593 microns, and a
conversion rate of 2.77 $\times\ 10^{-6}$ Jy ADU$^{-1}$ sec$^{-1}$.
Absolute photometric errors are dominated by the limitations of the 
PSF model and the zero point uncertainties of about 0.1 mag. 
The foreground Galactic extinction in the direction of PG~1115+080
is only $E(B-V)=0.041$ for $R_V=3.1$ (Schlegel, Finkbeiner \& Davis 1998),
hence, we applied no corrections to the numbers in Table 2.

The ratios of the quasar fluxes show little variation with wavelength,
and most of the observed variations are consistent with the level of
temporal variations seen by Schechter et al. (1997). In particular,
the curious flux ratio of the close A1/A2 pair is approximately $0.65$, 
independent of wavelength. If we fit the variations 
with wavelength of the flux ratios
with an extinction model using an $R_V=3.1$ Cardelli et al. (1989) 
extinction curve in the lens galaxy, we find that the differential 
extinction between the images is $|\Delta E(B-V)| \ltorder 0.02$ mag and the 
lens galaxy is virtually dust free. Simple lens models require an A1/A2 
flux ratio close to $0.9$ because the images are symmetrically arranged 
near a fold caustic (see Schneider, Ehlers \& Falco 1992). Since neither 
dust nor stellar microlensing is a viable explanation for the observed
ratio, we are presumably seeing the effects of a larger perturbation
in the gravitational potential produced by a globular cluster or a 
small satellite galaxy. Images near a caustic (like the A1/A2 pair) 
are particularly susceptible to magnification perturbations by 
irregularities of the potential that are too large to produce rapid
temporal variations (microlensing) but too small to appreciably alter 
the image positions or estimates of \Ho\ (Mao \& Schneider 1998).
 
Our NICMOS observations are the first to resolve the lens galaxy
clearly and allow detailed surface photometry models. We fit the lens
galaxy as an ellipsoidal de Vaucouleurs model simultaneously with the
quasar images using the Tiny Tim PSF model and a constant noise
variance. We find a good fit for an effective radius $R_e =
0\farcs59\pm0\farcs06$, surface brightness $\mu_e = 18.46 \pm
0.20~H$~mag arcsec$^{-2}$, and an extrapolated total magnitude 
$m_G = 16.26 \pm 0.05~H$
mag. The galaxy model is almost circular with a bound on the axis
ratio of $q > 0.93$ (at 1$\sigma$, $q>0.77$ at 3$\sigma$), and hence 
the position angle is unconstrained. For a redshift of $0.31$ the 
effective radius is $1.65\ h^{-1}$~kpc ($1.79\ h^{-1}$~kpc) for 
$\Omega_0 = 1$ ($\Omega_0 = 0$). We therefore confirm the lens
as a normal elliptical with a luminosity close to $L_{\ast}$. 

It is not straightforward to place the lens galaxy on the fundamental plane 
(FP) for ellipticals because the zero points depend on waveband, redshift, 
and possibly environment. The most direct comparison can be made with the
FP in the cluster Cl1358+62 at $z = 0.33$ (Kelson etal 1997), where no 
differential K-corrections, or evolutionary corrections are required.
We adopt $R_e = 2.36$~kpc (for $q_0 = 0.05$ and $H_0 = 75$ \kmm), and a
rest-frame $V$ band effective surface brightness of $\mu_e(V_{\rm rest})
=20.9$, which assumes that $f_\lambda(5500\AA)/f_\lambda(1.2\mu m)\approx 
1.7$ for a 9~Gyr old elliptical of solar luminosity (Vazdekis at al. 1996).
Taking $\sigma = 235$~km s$^{-1}$ from Table 3 (see also \S 2.3), the 
lens galaxy in PG1115 has parameters very similar to the FP defined
by the galaxies in Cl1358+62: it differs by less than 0.05 in $\log R_e$
and in $1.24\log \sigma - 0.33(\mu_e-25)$.

The lens models discussed in \S2.3 determine the mass of the system
projected inside the ring radius ($1\farcs15$) with internal uncertainties 
of only a few percent. This mass is the sum of the galaxy and group
masses, so the mass of the galaxy alone depends on the nature of the
group.  In models with a singular isothermal sphere (SIS) group, the
galaxy mass is $1.24\times10^{11}\ h^{-1}\ M_\odot$, while in models
with a point mass group the galaxy mass is $1.39\times10^{11}\ h^{-1}\
M_\odot$. We can combine the mass with the photometry to compute a
mass-to-light ratio.  The galaxy has an $H$ magnitude of 16.60 inside
$1\farcs15$. With models of the spectral evolution of elliptical
galaxies we can compute $K$ and evolutionary corrections and transform
to rest $V$ or $I$ magnitudes. Unfortunately, different spectral
evolution models do not completely agree on the extrapolation. The
models of Poggianti (1997) give a $V$-band magnitude of $M_V = -20.03
+ 5 \log h$ (for $\Omega = 1$), while the models of Bruzual \& Charlot
(1993) give $-20.35 + 5 \log h$. The main difference between the
spectral evolution models is the $z = 0$ colors. Poggianti sets the
$z = 0$ colors to be the colors of a galaxy at an age of 15 Gyr, while
our use of the Bruzual \& Charlot models has elliptical galaxies forming
at $z = 15$ for a present age of 12.8 Gyr (see Keeton, Kochanek \& Falco
1998). The color differences are smaller if we extrapolate only to
$I$ band, in which case the Poggianti (Bruzual \& Charlot) models give
$M_I = -21.46\ (-21.57) + 5 \log h$.

The lens models with an SIS
group then give $I$ band mass-to-light ratios of $(M/L)_I = 8.2$ (7.4),
while the lens models with a point mass group give $(M/L)_I = 9.2$ (8.3).
Adopting a $B-I$ color for an old (10~Gyr), solar-metallicity population,
these correspond to a more traditional $B$ band mass-to-light ratio
of $(M/L)_B = 14.2\ (12.9)$, and $(M/L)_B = 15.9\ (14.4)$, respectively.
These values are higher than expected from constant mass-to-light
stellar dynamical models (e.g. van der Marel 1991), suggesting that
we need dark matter, but this conclusion is not robust given the 
uncertainties in the spectral extrapolation.

\subsection{Lens Models and the Hubble Constant}

We fitted the quasar and lens galaxy positions, adopting 3~mas uncertainties,
and the quasar fluxes (20\% uncertainties) using the methods and models of
Keeton \& Kochanek (1997) and Courbin et al. (1997). We modeled the lens 
galaxy alternatively as a singular isothermal ellipsoid, an ellipsoidal 
Hubble law with a core radius of $0\farcs2$ ($\approx 0.56 h^{-1}$ kpc), 
and an ellipsoidal de Vaucouleurs law combined with a group modeled either 
as a singular isothermal sphere or a point mass. The de~Vaucouleurs model 
was constrained to match the observed profile of the lens galaxy, with 
an effective radius $R_e = 0\farcs6$, ellipticity $<0.07$ (1$\sigma$),
and an unconstrained position angle; in all other models, the ellipticity 
and position angle were unconstrained. All models led to a well defined
group location, a nearly circular lens galaxy even when the ellipticity 
was unconstrained, and a time delay ratio of $r_{ABC} \equiv
\Delta\tau_{AC}/\Delta\tau_{BA} \simeq 1.3 $. The predicted group 
location is near the luminosity-weighted centroid of the four bright 
group members (Kundi\'c et al. 1997), and the delay ratio is consistent 
with the measured $1.13^{+0.18}_{-0.17}$ (\cite{bar97}). Figure 3 shows 
the model group position in relation to the positions of the group galaxies 
and their centroid. Table 3 lists the parameters of all the variants of our
models. Lens positions are not listed because in all models they matched 
the observed galaxy position within 1 mas. We did not use explicitly the
constraints provided by the Einstein ring in our attempts to estimate 
model parameters, because it is not sufficiently bright. However, we 
used it to evaluate properties of models constrained by the point images. 
Figure 1c shows the NICMOS image after the quasar point sources and the
lens galaxy have been modeled and subtracted, clearly revealing the 
Einstein ring due to lensed light from the quasar host galaxy. 

All five model families produced acceptable, and statistically
indistinguishable, fits to the data, with a $\chi^2 \simeq 3$ (for 
$N_{dof}=1$) dominated by the poor fit to the anomalous A1/A2 flux ratio.
We note, however, that the variations in $\chi^2$ that yield the parameter
errors in Table 3 come equally from all terms and are not dominated
by the flux $\chi^2$. If the lens has dark matter and can be modeled 
as a singular isothermal ellipsoid, we find $H_0=44\pm4$ km s$^{-1}$ 
Mpc$^{-1}$; if the lens has a constant mass-to-light ratio, this value 
increases to $H_0 = 65\pm5$ km s$^{-1}$ Mpc$^{-1}$. 
If we use a point mass representation of the group, the 
value of $H_0$ rises by approximately 10\%. Because the four galaxies 
are group members, it is plausible that their halos have been tidally 
stripped to become part of the mean group halo. We fitted the lens 
using the ellipsoid $\rho \propto 1/r^2 (a^2+r^2)$ where $r^2$
is the ellipsoidal coordinate, to see how $H_0$ depends on the halo
truncation radius $a$. Inside $a$ the model has a flat rotation curve
and outside $a$ it becomes Keplerian. The model is similar to a
Jaffe (1983) model, but its lensing potential is analytic (see Keeton
\& Kochanek 1998). As expected, the value of $H_0$ does not change
significantly until the truncation radius is comparable to the ring
radius, with $H_0=47$, $49$, $53$, $56$ km s$^{-1}$ Mpc$^{-1}$ for
$a=10\arcsec$, $5\arcsec$, $3\arcsec$ and $2\arcsec$ ($27h^{-1}$, 
$14h^{-1}$, $8h^{-1}$ and $5h^{-1}$ kpc for $\Omega_0=1$), and we 
cannot distinguish between the models. 

Current evidence indicates that early-type galaxies possess 
dark matter such that the overall mass profile produces a nearly flat
rotation curve. It is found in stellar dynamical models (e.g. Rix et 
al. 1997), X-ray ellipticals (Fabbiano 1989), and gravitational lensing 
(Maoz \& Rix 1993; Kochanek 1995; Grogin \& Narayan 1996).  In the case 
of PG~1115+080, we consider the possibly high mass-to-light ratio as
supporting evidence for dark matter associated wirth the lens. However, 
a group member may have its halo tidally stripped and merged into the mean 
group halo. In PG~1115+080, however, the stripping scale for which our 
models yield $\Ho \geq 65$\kmm\ is extraordinarily small, particularly
since the velocity dispersion of the lens galaxy is comparable to that of 
the group. Moreover, models of the first lens with a time delay measurement,
Q~0957+561, require a nearly isothermal mass distribution on the same scales 
even though the galaxy is part of a small cluster rather than a small group
(Grogin \& Narayan 1996).

Note that we quote results only for an $\Omega_0=1$ cosmological model. For 
a low density, open universe with $\Omega_0=0.1$, the Hubble constant would
increase by 7.3\% and for a flat model dominated by a cosmological constant
($\lambda_0=0.8$) the Hubble constant would increase by 4.3\%.  Fluctuations
in the mean line of sight density produced by the non-linear power spectrum
of density fluctuations normalized by the local cluster abundance can produce 
an rms variance of 6\% (Seljak 1994; Barkana 1996), with the Hubble constant
increasing if there is a void in front of the lens. Most large perturbations
produced by the non-linear power spectrum should be positive and associated 
with visible perturbing objects in the WFPC1 images
(e.g., Keeton, Kochanek \& Seljak 1997).  

\subsection{The Nearby Galaxy Group}

The galaxy  group is an essential component of models for PG~1115+080 (Keeton 
et al. 1997; Schechter et al. 1997). In fact, some source of additional tidal
perturbations appears to be a ubiquitous requirement for good fits to all
well-constrained gravitational lenses (Keeton et al. 1997). The roundness 
of the lens galaxy means that in any constant $M/L$ model the astigmatism 
of the lens is almost entirely due to the group. Although we neither expect 
nor observe a strong correlation between the axis ratios of the light and 
the mass (see Keeton, Kochanek \& Falco 1998), even our dark matter models
predict that the lens galaxy mass distribution is nearly round. 

We have used the unpublished Westphal image to identify 9 new galaxies
in the field, within a radius of about 100 $h^{-1}$ kpc of the
previously known galaxies. Table 4 lists the positions and magnitudes
for these new galaxies, along with 3 galaxies that are detected on the
NICMOS frame but not with WFPC1. The offsets were determined using the
STSDAS task ``metric.'' Table 4 also includes 
comments on the morphology, although the aberrated HST images allow 
no more than a crude characterization of these galaxies. Photometry 
of the nearby galaxies is obtained using large aperture sizes that 
include most of their light. No ``deblending'' was performed for 
the bright, previously known companions G1, G2, and G3. At $z \approx  
0.3$ the sensitivity limit of the WFPC1 data reaches down to objects 
similar to the LMC, at $M_V \sim -16$. As in the Local Group, there are 
two luminous spirals (G2 and G3), one of which has two dwarf satellite 
companions. Most of the other galaxies appear to be early type, although 
there is one nucleated dwarf. The luminosity weighted centroid of all 
detected galaxies is $d = 18\farcs7$ at a position angle of $-117^\circ$
(G12 has been excluded because it may not be a galaxy). This can be 
compared to $d = 13\farcs5$ at position angle $-115^\circ$ for the 
centroid C4 of the main lens galaxy plus G1, G2, and G3. In our 
refined lens models, the group location in the lens models is well 
constrained and lies near the centroid C4, with small positional 
shifts depending on the assumed model (see Figure 3). 

With an early-type fraction of 1/3 to 1/2 in this group, we would
expect a velocity dispersion in the range 250-400 km s$^{-1}$
(\cite{zab97}), consistent with the mass required by the successful
lens models and the value measured by Kundi\'c et al. (1997) of
$270\pm70$ km s$^{-1}$. Among the isothermal group models, the group
has a velocity dispersion ranging from $320\kms$ for the singular
isothermal ellipsoidal lens to $430\kms$ for the de Vaucouleurs lens
galaxy. The mass inside a cylinder of radius equal to the distance
from the group center to the lens galaxy ranges from $2.1h^{-1} \times
10^{12} M_\odot$ to $3.5h^{-1} \times 10^{12} M_\odot$. The light from
the group in the same cylinder is simply the sum of the lens galaxy,
G1, G2, and G3 -- for a total of $1.7 \times 10^{10}\ \lsolar$ or
about $3 L_*$. This integral has converged with these four galaxies
and does not depend on the group membership of other faint galaxies in
the WFPC1 images. We derive a mass-to-light ratio of 40-90 for the $B$
band within a radius of $\sim 39 h^{-1}$ kpc, the distance C4-G.
These values are consistent with the local estimates for small groups
of Ramella, Pisani \& Geller (1997), because their estimates used a
median radius of $\sim 400$ kpc which would include ten times more mass
but no additional light.

\subsection{The Quasar Host Galaxy}

In addition to the quasar images and the lens galaxy, we also clearly
detect an Einstein ring formed by the lensed image of the quasar host
galaxy.\footnote{The ring is also seen in deep WFPC2 I band images 
(Schechter 1998, private communication).} Figure 1c shows the Einstein ring, 
where residuals remain due to imperfect subtraction of the strong and
undersampled quasar images. The flux in the ring is well above the 
background ($10\sigma$ to $70\sigma$ at a 
rms noise of 0.001-0.002 ADU pixel$^{-1}$ sec$^{-1}$). If we break the 
ring up into polygonal sectors between quasar images A2 and C (\#1), C 
and B (\#2), and B and A1 (\#3), we find mean $H$ band surface brightnesses
of 19.43$\pm$0.07 mag arcsec$^{-2}$ (\#1), 20.06$\pm$0.06 mag arcsec$^{-2}$
(\#2), and 18.92$\pm$0.05 mag arcsec$^{-2}$ (\#3). The point source
contamination in the aberrated WFPC1 images is too severe to derive
a color or a limit on the color of the ring.

We used the lens models fitted in \S2.3 to produce models of the ring
by adding a lensed exponential disk quasar host galaxy to our 
photometric model. 
Figure 1d shows the image that our de Vaucouleurs
main lens galaxy and isothermal group would form, after convolution with
the appropriate Tiny Tim PSF. 
The models reproduce the structure of the ring
well, but the details of galaxy parameters are sensitive to the large
residuals (compared to the ring) created by the mismatch between the 
model PSF and the quasar images. Detailed modeling of the ring would 
require a PSF model that matches the observations very accurately. The 
best fit intrinsic magnitude of the host galaxy is 20.6 $H$ mag, compared to 
18.7 $H$ mag for the quasar, in the SIE/SIS model. Using K-corrections 
and an evolutionary correction for a passively evolving old stellar 
population, we compute present-day absolute magnitudes of $M_B=-23.6 + 
5 \ {\rm log} h$ ($\Omega = 0.1$) for the quasar and $M_H = -23.4$, 
i.e. $L_*$, for the galaxy. Hence, given its host's luminosity, the 
quasar is at the maximum luminosity observed in a sample of low-redshift
quasars (McLeod \& Rieke 1995).

%

Since the models fitted to the quasars and the lens galaxy position 
are degenerate with respect to the value of the Hubble constant, we 
explored whether improved data on the ring could distinguish between
models. The primary difference between the constant $M/L$ and flat 
rotation curve models lies in the radial magnification profile near the 
Einstein ring. We took the best fit ring found using the constant $M/L$ 
galaxy plus SIS group lens model, and then attempted to fit the model 
ring using the SIE galaxy plus SIS group lens model while adjusting 
the source galaxy parameters to find the best fit. The largest 
differences are concentrated near the center of the quasar host galaxy and
lie under the quasar images, but significant residuals appear in the 
regions away from the quasar images. The best-fit source for the SIE
model is too broad to fit the ``de~Vaucouleurs ring" regardless
of the host galaxy's scale length, leaving residuals at the level 
of $\sim 0.01$ ADU pixel$^{-1}$ sec$^{-1}$, or a surface brightness
of $20.8$ $H$ mag arcsec$^{-2}$ in the regions away from
the quasar images. Hence, images with accurately matched PSFs and 
with significantly reduced contrast between the quasar and host 
galaxy images could directly discriminate between lens models of differing 
radial mass profile.

\section{DISCUSSION}

New infrared data on PG~1115+080 affirms multiple-component gravitational
lens systems as powerful cosmological tools. The major puzzle remaining in 
the PG~1115+080 system is the anomalous A1/A2 flux ratio. Our observations 
rule out differential extinction as an explanation, and microlensing is 
ruled out by its lack of variability. Since a flux ratio near 0.9 is a 
generic feature of the large scale potential near a fold caustic, only a
potential perturbation intermediate between that produced by isolated 
stars (microlensing) and by the overall galaxy can explain the flux ratio.
The potential of PG~1115+080 must be perturbed either by a satellite galaxy 
or a globular cluster. Mao \& Schneider (1998) showed that such perturbations
alter the time delay -- and so the inferred value of the Hubble constant --
fortunately by no more than 2-3\%.

Our improved astrometry greatly reduces some of the degeneracies in early 
models of the system. The group position is now well constrained and located
near the luminosity centroid of the four bright group galaxies, and the lens
galaxy is constrained to be nearly circular. Unfortunately, the degeneracies
in the $H_0$ estimate have been exacerbated because with the revised 
astrometry the models no longer favor dark matter models over constant 
$M/L$ models for the main lens galaxy. Because all 4 images are located
at nearly the same radial distance from the center of the lens galaxy, we
do not expect the models to be sensitive to the radial mass profile of the 
lens galaxy (Kochanek 1991; Wambsganss \& Pa\'czynski 1994). We find \Ho\ 
ranging from $44\pm4$ \kmm\ if the lens galaxy is modeled as a singular
isothermal ellipsoid and the group as a singular isothermal sphere, to 
$65\pm5$ ($72\pm5$) \kmm\ for $\Omega_0 = 1$ ($0.1$) if the lens galaxy 
has a constant $M/L$. Note that we find evidence for dark matter in our
high value of $M/L$.
A model with an adjustable truncation radius shows that the halo must be
truncated on scales comparable to the ring diameter for $H_0$ to exceed 
$60$ \kmm. Such a halo seems smaller than physically plausible given 
that the velocity dispersions of the group and the lens galaxy are 
comparable.  

Further progress in reducing the uncertainties depends on improving
the time delay measurements and on making more detailed studies of the
Einstein ring formed by the quasar host galaxy. First, for any given
mass profile, most of the current errors 
in \Ho\ are due to the uncertainties
in the time delays. Second, all the best fit models predict time delay
ratios near $r_{ABC}=1.3$, consistent with the current measurement of
$1.13 \pm 0.18$. If nothing else, a more accurate measurement of the
delay ratio than is now available would be a powerful test of
the models. For any given lens mass profile, the ratio constraint
would further reduce the parameter space for the position and mass of
the group, or could be used to constrain more complicated models 
(e.g., Saha \& Williams 1997). Deep new observations to determine the 
surface brightness of the ring accurately, combined with direct 
measurement of the point spread function at the time of the observations 
would probably permit us directly to break the degeneracy of the models. 
Only NICMOS, however, has the ability to make these difficult 
observations within a decade.

No single technique or observation can tie down the Hubble constant --
the long history of unrecognized or underestimated systematic errors
in this subject encourages humility. Nevertheless, PG~1115+080
demonstrates the potential of the gravitational lens approach. With
recent reductions in age estimates of the oldest globular clusters
(Chaboyer et al. 1998), and the likelihood that the mass density of
the universe is lower than the Einstein-de Sitter case (e.g.,
Garnavich et al. 1998), the possibility of an age conflict in the big
bang model has receded. The modeling of PG~1115+080 gives a plausible
upper bound on the Hubble constant if we accept that the group is not
a point mass and that the lens galaxy is unlikely to have a mass
distribution that is more concentrated than its light
distribution. This bound is $H_0 < 67$ ($72$) \kmm\ for $\Omega_0 = 1$
($0.1$). The most recent result of the HST Extragalactic Distance
Scale Key Project is $H_0 = 72 \pm5$ (random) $\pm12$ (systematic) km
s$^{-1}$ Mpc$^{-1}$ (Madore et al. 1998). Our upper limit is inconsistent
with the upper end of the range from the Key Project, although it is
consistent with the lower end of the range. There appears to be 
satisfactory concordance among the basic parameters of the big 
bang model, and between direct and indirect measures of the distance
scale. Gravitational lenses can be expected to play an increasing role
as versatile cosmological tools.

\vskip 1truecm

\acknowledgments

We would like to thank many people in the NICMOS team who provided advice
and technical information. In particular, Marcia Rieke and Dean Hines helped
with photometry issues and Glenn Schneider, Ray Lucas and the STScI Help 
Desk helped with astrometric issues. CDI acknowledges support from NSF under
grant AST-9320715. CSK and CRK acknowledge financial support from NASA under 
ATP grant NAG5-4062, and from NSF under grant AST-9401722. Support for this 
work was provided by NASA through grant number GO-7495 from the Space 
Telescope Science Institute, which is operated by the Association of
Universities for Research in Astronomy, Inc., under NASA contract 
NAS5-26555.

\clearpage

\begin{deluxetable}{lcccccccc}
\footnotesize
\tablecaption{Astrometry of the PG 1115$+$080 System. \label{tbl-1}}
\tablewidth{0pt}
\tablehead{
\colhead{} & \multicolumn{2}{c}{HST/NICMOS\tablenotemark{a}} &
\multicolumn{2}{c}{HST/WFPC1\tablenotemark{b}} & 
\multicolumn{2}{c}{HST/WFPC1\tablenotemark{c}} &
\multicolumn{2}{c}{NOT\tablenotemark{d}} \\
\colhead{Image} & 
\colhead{$\Delta$(RA)} & \colhead{$\Delta$(Dec)} &
\colhead{$\Delta$(RA)} & \colhead{$\Delta$(Dec)} & 
\colhead{$\Delta$(RA)} & \colhead{$\Delta$(Dec)} &
\colhead{$\Delta$(RA)} & \colhead{$\Delta$(Dec)} \\
\colhead{} & \colhead{(\arcsec)} & \colhead{(\arcsec)} & 
\colhead{(\arcsec)} & \colhead{(\arcsec)} & 
\colhead{(\arcsec)} & \colhead{(\arcsec)} &
\colhead{(\arcsec)} & \colhead{(\arcsec)} \nl
}
\startdata
A1   & $+$1.328 & $-$2.037 & $+$1.318 & $-$2.032 & $+$1.313 & $-$2.031 & $+$1.291  & $-$2.028 \nl
A2   & $+$1.478 & $-$1.576 & $+$1.468 & $-$1.578 & $+$1.463 & $-$1.577 & $+$1.445  & $-$1.578 \nl
B    & $-$0.341 & $-$1.960 & $-$0.350 & $-$1.956 & $-$0.346 & $-$1.956 & $-$0.364  & $-$1.940 \nl
C    &$\equiv$0 &$\equiv$0 &$\equiv$0 &$\equiv$0 &$\equiv$0 &$\equiv$0 &$\equiv$0  &$\equiv$0 \nl
Lens & $+$0.382 & $-$1.344 & $+$0.386 & $-$1.362 & \dots    & \dots    & $+$0.332 & $-$1.339 \nl
\tablenotetext{a}{HST/NICMOS F160W image presented in this paper. The rms
internal error on each measurement is 0\farcs002. Plate scales are
$0\farcs076030$ per pixel in X and $0\farcs0.075344$ per pixel in Y with 
an orientation of $68\pdeg760$. }
\tablenotetext{b}{HST/WFPC1 F785LP image W93, previously 
unpublished. The plate scale is $0\farcs04404$ with an orientation of
$123\pdeg087$. }
\tablenotetext{c}{HST/WFPC1 F785LP, new measurements of an image published
by Kristian et al. (1993). }
\tablenotetext{d}{For comparison, Nordic Optical Telescope $I$ band results 
of Courbin et al. (1997) using an astrometric solution adapted to positions
measured by Kristian et al. (1993). }
\enddata             
\end{deluxetable}

\clearpage

\begin{deluxetable}{cccccccccc}
\footnotesize
\tablecaption{Photometry of the PG 1115$+$080 System.\tablenotemark{a}
   \label{tbl-2}}
\tablewidth{0pt}
\tablehead{
\colhead{} & \colhead{NICMOS\tablenotemark{b}}
& \multicolumn{2}{c}{HST/WFPC1\tablenotemark{c}}
& \multicolumn{2}{c}{HST/WFPC1\tablenotemark{d}} 
& \colhead{NOT\tablenotemark{e}}
& \multicolumn{3}{c}{CFHT\tablenotemark{f}}  
\\
\colhead{Image} & \colhead{$H$} 
& \colhead{$I$} & \colhead{$V$} & \colhead{$I$} & \colhead{$V$} 
& \colhead{$I$} & \colhead{$R$} & \colhead{$V$} & \colhead{$B$} 
\\
\colhead{} & \colhead{(mag)}  & \colhead{(mag)}  & \colhead{(mag)}  
& \colhead{(mag)}  & \colhead{(mag)}  & \colhead{(mag)}  
& \colhead{(mag)}  & \colhead{(mag)}  & \colhead{(mag)} \nl
}
\startdata
A1   & 15.44$\pm$0.02 & 16.12 & 16.90 & 16.19 & 16.81 & 16.34 & 
16.71 & 16.99 & 17.48 \nl
A2   & 15.92$\pm$0.03 & 16.51 & 17.35 & 16.56 & 17.32 & 16.75 &
16.95 & 17.27 & 17.74 \nl
B    & 17.37$\pm$0.04 & 18.08 & 18.87 & 18.05 & 18.73 & 18.30 &
18.46 & 18.74 & 19.19 \nl
C    & 16.92$\pm$0.03 & 17.58 & 18.37 & 17.60 & 18.40 & 17.82 &
17.97 & 18.26 & 18.71 \nl
Lens & 16.26$\pm$0.10 & 18.40 & \dots & 18.55 & \dots & 19.60 &
20.20 & 20.90 & \dots \nl
\tablenotetext{a}{Magnitudes are not corrected for the estimated foreground
extinction of $E(B-V)=0.041$ mag estimated by Schlegel, Finkbeiner \& Davis
(1998). }
\tablenotetext{b}{H(F160W) band data from this paper. The zero point is
uncertain by 0.1 mag, and the total lens magnitude is from a model fit. }
\tablenotetext{c}{I(F785LP) and V(F555W) band data image from Kristian
et al. (1993). The zero points are from Holtzman et al. (1991). }
\tablenotetext{d}{I(F785LP) and V(F555W) band data image from unpublished
W93 image. The zero points are from Holtzman et al. (1991). }
\tablenotetext{e}{I band data from Courbin et al. (1997). The lens magnitude 
is measured in a 0\farcs9 aperture, and no zero point error is quoted. }
\tablenotetext{f}{BVR band data from Christian et al. (1987). The lens 
magnitude is measured in a 1\farcs6 aperture, with a zero point accurate 
to 0.05 mag. }
\enddata             
\end{deluxetable}

\clearpage

\voffset -1truecm
\begin{deluxetable}{llllll}
\footnotesize
\tablecaption{Lens Model Results \label{tbl-3}}
\tablehead{& SIE\tablenotemark{a} / SIS & Hub\tablenotemark{b} / SIS &
	de Vauc\tablenotemark{c} / SIS & SIE\tablenotemark{a} / Pt & Hub\tablenotemark{b} / Pt}
\startdata
{\rm \qquad Galaxy\/}\tablenotemark{d} & & & & & \\
$b$ ($''$)
	& $1.04\pm0.01$ & $1.03\pm0.01$ & $1.03\pm0.01$ & $1.15\pm0.01$ & $1.15\pm0.01$ \\
$M$ ($10^{11} h^{-1} M_\odot$)
	& $1.25\pm0.02$ & $1.24\pm0.01$ & $1.24\pm0.01$ & $1.39\pm0.01$ & $1.38\pm0.01$ \\
$\sigma$ (km/s)
	& $230\pm1$ & --- & --- & $243\pm1$ & --- \\
$q$
	& $0.96\pm0.03$ & $0.96\pm0.04$ & $0.96\pm0.04$ & $0.95\pm0.02$ & $0.96\pm0.02$ \\
PA ($^\circ$)
	& $46_{-34}^{+16}$ & $61_{-44}^{+17}$ & $68$\tablenotemark{d}
	& $28_{-25}^{+20}$ & $30_{-25}^{+33}$ \\
\tableline
{\rm \qquad Group\/}\tablenotemark{e} & & & & & \\
$b$ ($''$)
	& $2.0\pm0.4$ & $3.1\pm0.4$ & $3.6\pm0.4$ & $4.3\pm0.7$ & $5.2\pm0.6$ \\
$M$ ($10^{11} h^{-1} M_\odot$)
	& $20.8\pm5.0$ & $34.9\pm6.7$ & $42.5\pm7.2$ & $19.7\pm6.3$ & $27.9\pm7.0$ \\
$\sigma$ (km/s)
	& $320\pm32$ & $397\pm26$ & $428\pm24$ & --- & --- \\
$d_{grp}$ ($''$)
	& $10.0\pm1.3$ & $10.8\pm1.5$ & $11.3\pm1.3$ & $12.8\pm1.6$ & $12.6\pm1.6$ \\
$\theta_{grp}$ ($^\circ$)
	& $-113\pm2$ & $-116\pm1$ & $-116\pm1$ & $-112\pm2$ & $-114\pm1$ \\
\tableline
$H_0$ (km s$^{-1}$ Mpc$^{-1}$)
	& $44\pm4$ & $61\pm5$ & $65\pm5$ & $47\pm4$ & $68\pm6$ \\
\tableline
{\rm \qquad Best-fit Model\/}\tablenotemark{f} & & & & & \\
Source $\Delta\alpha$ ($''$) & $-1.79$ & $-2.73$ & $-3.16$ & $-1.33$ & $-1.88$ \\
Source $\Delta\delta$ ($''$) & $-0.69$ & $-1.18$ & $-1.41$ & $-0.44$ & $-0.71$ \\
Source Flux                  & $ 0.17$ & $ 0.33$ & $ 0.39$ & $ 0.20$ & $ 0.42$ \\
Total Magnification          & $ 46  $ & $ 24  $ & $ 20  $ & $ 40  $ & $ 19  $ \\
$\chi^2$ (positions)         & $ 0.03$ & $ 0.05$ & $ 0.04$ & $ 0.10$ & $ 0.05$ \\
$\chi^2$ (fluxes)            & $ 3.58$ & $ 3.06$ & $ 3.06$ & $ 4.36$ & $ 4.14$ \\
$\chi^2$ (galaxy)            & $ 0.02$ & $ 0.05$ & $ 0.26$ & $ 0.12$ & $ 0.08$ \\
$\chi^2$ (total)             & $ 3.63$ & $ 3.16$ & $ 3.36$ & $ 4.57$ & $ 4.27$ \\
\enddata
\tablenotetext{a}{Singular isothermal ellipsoid galaxies with an unconstrained
ellipticity and position angle.}
\tablenotetext{b}{Ellipsoidal Hubble model galaxies with a fixed core radius
of $0\farcs2 = 0.56 h^{-1}$ kpc, and an unconstrained ellipticity and position
angle.}
\tablenotetext{c}{Ellipsoidal de Vaucouleurs model galaxies constrained to
match the observed profile of the lens galaxy, with an effective radius
$R_e = 0\farcs6$, an ellipticity $< 0.07$ ($1\sigma$), and an unconstrained
position angle. The model PA can take any value at the $\Delta\chi^2 = 1$
level.}
\tablenotetext{d}{The best-fit values and $\Delta\chi^2 = 1$ error bars result
from the variation of the lens galaxy's critical radius $b$, position, axis
ratio $q$, and position angle. No PA error bar is quoted for the de Vaucouleurs
galaxy model because the PA can take any value at the $\Delta\chi^2 = 1$ level.
The mass $M$ inside the ring radius $1\farcs15$, and $\sigma$ for isothermal
galaxies is given.}
\tablenotetext{e}{We varied the group critical radius $b$, and position in
polar coordinates $(d_{grp},\theta_{grp})$ centered on the lens galaxy.
We give the mass $M$ inside a cylinder of radius equal to $d_{grp}$, and
the velocity dispersion $\sigma$ for SIS groups.}
\tablenotetext{f}{Parameters for the best-fit models, including the source
position (relative to the lens galaxy), the source flux (relative to image
$A_1$), the total magnification, and the contribution to the $\chi^2$ from
the image positions, the fluxes, and the galaxy position.}
\end{deluxetable}

\clearpage
\begin{deluxetable}{cccccccrrl}
\footnotesize
\tablecaption{Nearby Galaxies in the PG~1115+080 Field \label{tabl-4}}
\tablewidth{500pt}
\tablehead{
\colhead{Name} & \multicolumn{2}{c}{F785LP\tablenotemark{a}} &
\multicolumn{2}{c}{F555W} & \multicolumn{2}{c}{F160W\tablenotemark{b}} &
\colhead{RA\tablenotemark{c}} & \colhead{Dec} 
& \colhead{Comments\tablenotemark{d}} 
\\
\colhead{} & \colhead{$I$}&  \colhead{$\sigma(I)$} & \colhead{$V$} &
\colhead{$\sigma(V)$} & \colhead{$H$} & \colhead{$\sigma(H)$} & 
\multicolumn{2}{c}{Offset} & 
\\
\colhead{} & \multicolumn{2}{c}{(mag)} & \multicolumn{2}{c}{(mag)} &
\multicolumn{2}{c}{(mag)} & \multicolumn{2}{c}{(arcsec)} & \colhead{} 
\nl
}
\startdata
Lens\tablenotemark{e} & 18.55 & 0.50 & \dots & \dots & 16.26 & 0.10 & 
0.382 & $-$1.344 & Elliptical, $z=0.310$\tablenotemark{f} 
\nl                          
G1  & 17.85 & 0.01 & 20.40 & 0.03 & \dots & \dots & $-$20.139 & $-$12.347 &
Sp/S0, $z=0.310$ \nl
G2  & 18.73 & 0.04 & 21.05 & 0.06 & \dots & \dots & $-$11.547 & $-$2.167 &
Sp, face-on, $z=0.312$ \nl
G3  & 19.44 & 0.02 & 21.69 & 0.07 & \dots & \dots & $-$13.607 & $-$13.516 &
Sp, edge-on, $z=0.309$ \nl
G4  & 19.88 & 0.04 & 22.46 & 0.12 & \dots & \dots & $-$61.151 & $-$19.406 &
Sp/S0? \nl
G5  & 20.63 & 0.06 & \dots & \dots & \dots & \dots & $-$17.588 & $-$40.630 &
E? \nl
G6  & 21.38 & 0.13 & \dots & \dots & \dots & \dots & $-$26.196 & $-$31.996 &
dE? \nl
G7  & 21.40 & 0.08 & \dots & \dots & \dots & \dots & $-$34.785 & $-$24.919 &
dIrr? \nl
G8  & 21.52 & 0.08 & \dots & \dots & \dots & \dots & $-$27.706 & $-$13.546 &
dE? \nl
G9  & 22.05 & 0.27 & \dots & \dots & \dots & \dots & $-$13.311 & $-$5.643 &
dw companion to G2 \nl
G10 & 22.09 & 0.09 & 23.56 & 0.10 & \dots & \dots & $-$41.203 & $-$3.078 &
dIrr? \nl
G11 & 22.98 & 0.34 & \dots & \dots & \dots & \dots & $-$10.957 & $-$4.892 &
dw companion to G2 \nl
G12 & 18.55 & 0.08 & \dots & \dots & \dots & \dots & 2.946 & $-$43.246 & K93 image (star?) \nl
G13 & \dots & \dots & \dots & \dots & 21.36 & 0.13 & 7.934 & $-$4.575 &
\nl
G14 & \dots & \dots & \dots & \dots & 20.94 & 0.07 & 3.494 & $-$7.962 &
\nl
G15 & \dots & \dots & \dots & \dots & 21.47 & 0.10 & 3.108 & $-$13.234 &
\nl
\enddata
\tablenotetext{a}{F785LP and F555W data derived from W93 images}
\tablenotetext{b}{F160W data from the NICMOS image presented in this paper}
\tablenotetext{c}{Offsets are calculated relative to component C of
PG~1115+080}
\tablenotetext{d}{Morphological information from W93 images, except where
otherwise indicated}
\tablenotetext{e}{Offset for the lens from our NICMOS image. The magnitude 
is our own estimate from fitting the point images and a de Vaucouleurs 
profile to the W93 I-band image}
\tablenotetext{f}{All redshifts are from Kundi\'c et al. (1997)}
\end{deluxetable}

\clearpage

\clearpage
\plotone{f1.eps}

\figcaption{NICMOS image of the PG~1115$+$080 system, taken with Camera 2
and the F160W filter. Each panel is $\sim 7\arcsec\times7\arcsec$, and 
is oriented with North at the top and East at the left.
(a) The sum of the dithered, flat-fielded exposures.
The four quasar images are separated by $\sim 1\farcs3$. (b) The same 
image after fitting and subtracting quasar point sources. Artifacts remain 
due to imperfect subtraction of the point spread function. The 
main lens galaxy is well resolved; an Einstein ring is perceptible. 
(c) The same image after subtracting the main lens galaxy, modeled 
as a de Vaucouleurs profile; a nearly complete
Einstein ring is now easily visible. (d) The Einstein 
ring that the de Vaucouleurs + SIS model predicts, after convolution with an
appropriate Tiny Tim PSF. Note that the ring is closed: there is 
very faint flux between images B and C.
}

\clearpage
\plotone{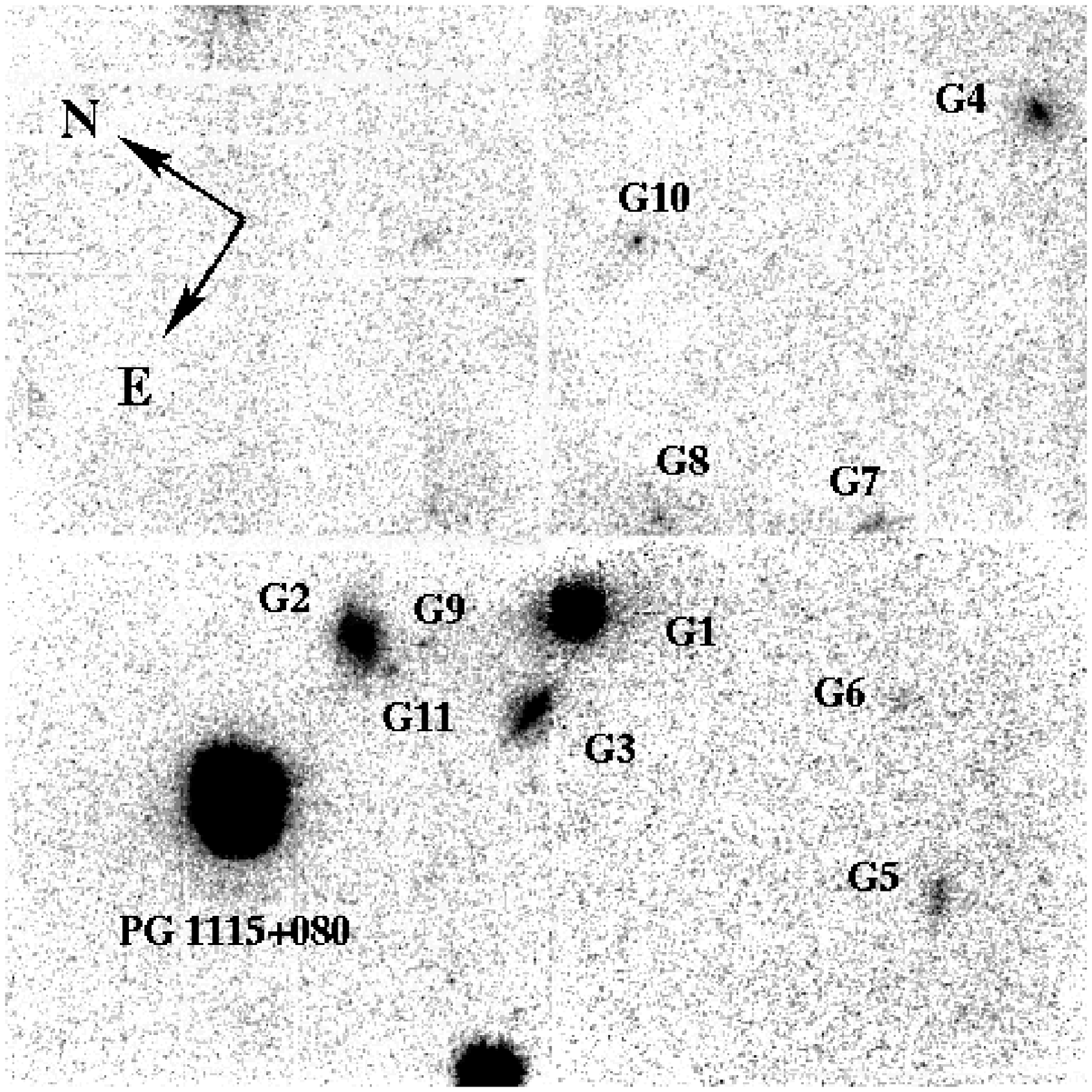}

\figcaption{Archival WFPC1 image of the PG~1115+080 system (W93), 
taken through the F785LP filter. We smoothed the image with a gaussian 
with $\sigma = 2$ pixels, to improve the contrast for the faintest features.
The multiply imaged quasar is saturated in this representation.
Nine galaxies near the lens system are listed in Table 4, along with 3
galaxies seen only on the NICMOS image. Although part of 
the bright object at the bottom of the image fell outside the WFPC1
field of view, it clearly appears to be a star.}

\clearpage
\plotone{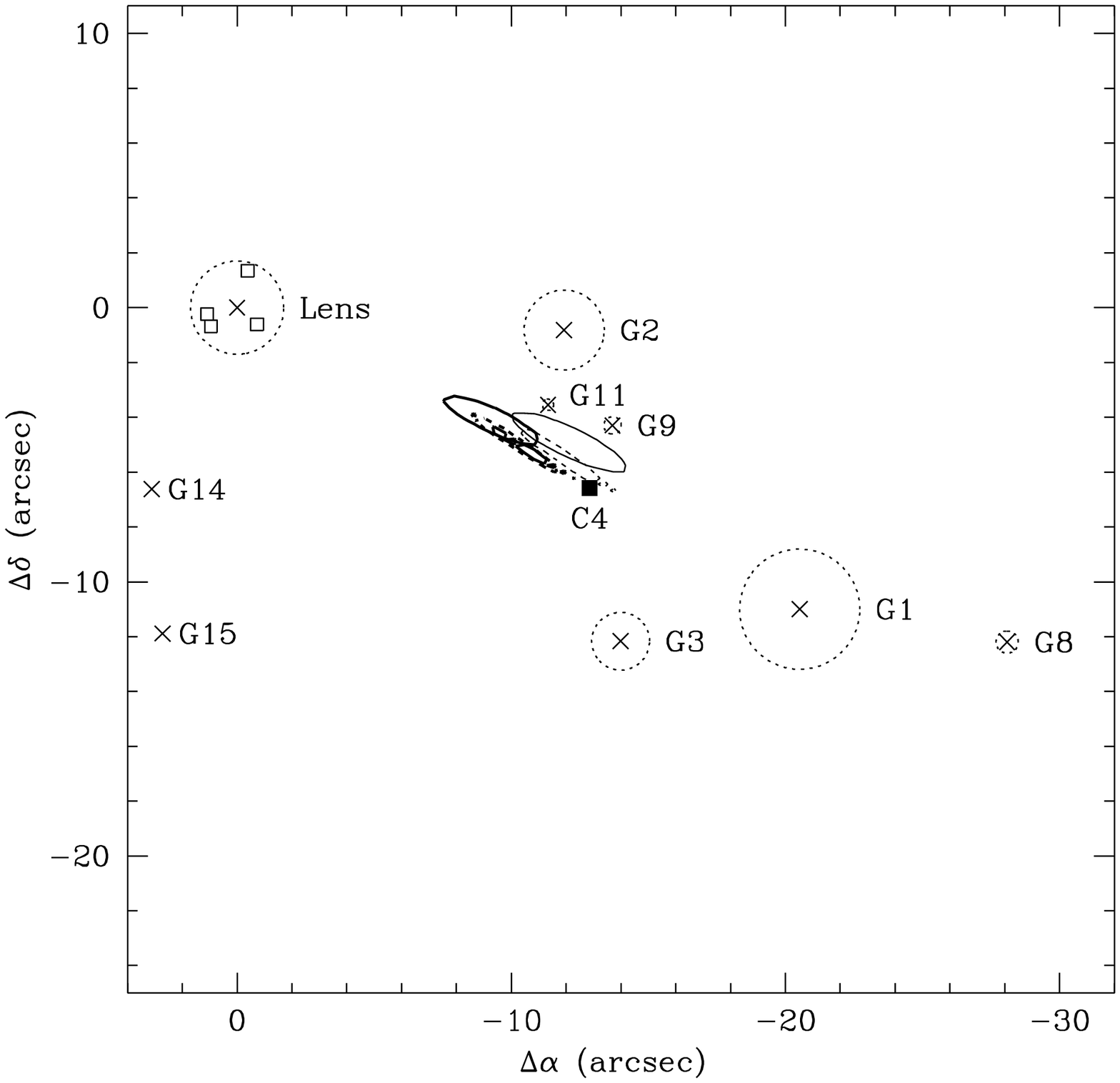}

\figcaption{The model group positions 
on the sky, shown as position error ellipses (bold solid: 
isothermal galaxy + SIS; light solid: isothermal galaxy + point mass; 
bold dotted: Hubble galaxy + SIS; light dotted: Hubble galaxy + point mass;
bold dashed: de Vaucouleurs galaxy + SIS). Also shown are the positions of 
the lensed images, the main lens galaxy, the group galaxies and the 
centroid C4 of the main lens galaxy plus G1, G2, and G3. The area of 
each dotted circle is proportional to the F785LP flux of the 
corresponding galaxy.
}
\newpage 


\clearpage
\begin{thebibliography}{}

\bibitem[Angonin-Willaime, Hammer, \& Rigaut 1993]{ang93} Angonin-Willaime,
	M.-C., Hammer, F., \& Rigaut, F. 1993, in Gravitational Lenses in the
	Universe, eds. J. Surdej et al. (Universite de Liege: Liege), p. 85
\bibitem[Barkana 1996]{bar96} Barkana, R. 1996, \apj, 468, 17
\bibitem[Barkana 1997]{bar97} Barkana, R. 1997, \apj, 489, 21
\bibitem[Bernstein et al. 1997]{} Bernstein, G., Fischer, P., Tyson, J. A. 
	\& Rhee, G. 1997, \apj, 483, 79
\bibitem[Bruzual \& Charlot 1993]{} Bruzual, G., \& Charlot, S. 1993, \apj, 
	405, 538
\bibitem[Cardelli et al. 1989]{car97} Cardelli, J., Clayton, G. C.
\& Mathis J. S., \apj, 1989, 345, 245
\bibitem[Chaboyer et al. 1998]{cha98} Chaboyer, B., Demarque, P., Kernian, 
	P.J., \& Krauss, L.M. 1998, \apj, 496, 96
\bibitem[Christian, Crabtree \& Waddell (1987)]{chr87} Christian, C.A., 
	Crabtree, D., \& Waddell, P. 1987, \apj, 312, 45
\bibitem[Corbettt et al. 1996]{cor96} Corbett, Browne, I.W.A., Wilkinson, P.N.,
	\& Patnaik, A.R. 1996, in Astrophysical Applications of Gravitational
	Lensing, eds. C.S. Kochanek and J.N. Hewitt 
        (Kluwer Academic Publishers: Dordrecht), p. 37
\bibitem[Courbin et al. 1997]{cou97} Courbin, F., Magain, P., Keeton, C.R.,
	Kochanek, C.S., Vanderriest, C., Jaunsen, A.O., \& Hjorth, J. 1997,
	\aap, 324, L1
\bibitem[Cox et al. 1997]{cox97} Cox, C., Ritchie, C., Bergeron, E., Mackenty,
	J., \& Noll, K. 1997, NICMOS Instrument Science Report 97-07, Space
	Telescope Science Institute
\bibitem[Fabbiano 1989]{fab89} Fabbiano, G. 1989, \araa, 27, 87
\bibitem[Fassnacht et al. 1996]{fas96} Fassnacht, C.D., Womble, D.S., 
	Neugebauer, G., Browne, I.W.A., Readhead, A.C.S., Matthews, K., 
	\& Pearson, T.J. 1996, \apj, 460, 103
\bibitem[Garnavich et al. 1998]{gar98} Garnavich, P., M., Schommer, R., 
	Schmidt, B., Jha, S., Challis, P., Filippenko, A.V., Riess, A.G. 
	\& Leonard, D.C. 1998, \apjl, 493, 53
\bibitem[Gilmozzi et al. 1995]{gil95} Gilmozzi, R., Ewald, S., \& Kinney, E.
	1995, WFPC2 Instrument Science Report 95-02, Space Telescope Science
	Institute
\bibitem[Gould \& Yanny 1994]{gou94} Gould, A., \& Yanny, B. 1994, \pasp, 
	106, 101
\bibitem[Grogin \& Narayan 1996]{gro96} Grogin, N.A., \& Narayan, R. 1996,
	\apj, 464, 92
\bibitem[Haarsma et al. 1997]{haa97} Haarsma, D.B., Hewitt, J.N., Leh\'ar, J.,
	\& Burke, B.F. 1997, \apj, 479, 102
\bibitem[Hege et al. 1981]{heg81} Hege, E.K., Hubbard, E.N., Strittmatter, 
	P.A., \& Worden, S.P. 1981, \apjl, 248, L1
\bibitem[Henry \& Heasley (1986)]{hen86} Henry, J.P., \& Heasley, J.N. 1986, 
	\nat, 321, 142
\bibitem[Holtzman et al. 1991]{hol91} Holtzman, J., Groth, E.J., Light, R.M.,
	Faber, S.M., et al. 1991, \apj, 369, L35
\bibitem[Jaffe 1983]{jaf83} Jaffe, W. 1983, \mnras, 202, 995
\bibitem[Jorgensen, Franx \& Kjaergaard 1996]{jor96} Jorgensen, I., Franx, M.,
	\& Kjaergaard, R. 1996, \mnras, 280, 167
\bibitem[Keeton \& Kochanek (1997)]{kee97} Keeton, C.R., \& Kochanek, C.S. 
	1997, \apj, 487, 42
\bibitem[Keeton \& Kochanek (1998)]{kk98} Keeton, C.R \& Kochanek, 
	C.S. 1998, 495, 157
\bibitem[Keeton, Kochanek, \& Falco (1998)]{kkf98} Keeton, C.R, Kochanek, 
	C.S., \& Falco, E. E. 1998, \apj\ in press (astro-ph/9708161)
\bibitem[Keeton, Kochanek \& Seljak 1997]{kef97} Keeton, C.R., Kochanek, C.S.,
	\& Seljak, U. 1997, \apj, 482, 604
\bibitem[Kelson et al. 1997]{kel97} Kelson, D.D., van Dokkum, P.G., Franx, M.,
	Illingworth, G.D., \& Fabricant, D. 1997, \apj, 478, L13
\bibitem[Kochanek 1995]{koc95} Kochanek, C.S. 1995, \apj, 453, 545
\bibitem[Krist \& Hook 1997]{kri97} 
	Krist, J. E. \& Hook, R. N. 1997, The Tiny Tim User's Guide, version
        4.4 (Baltimore: STScI)
\bibitem[Kristian et al. (1993)]{kri93} Kristian, J., Groth, E.J., Shaya, E.J.,
	et al. 1993, \aj, 106, 1330 (K93)
\bibitem[Kundi\'c et al. 1997a]{kun97a} Kundi\'c, T., Turner, E.L., Colley,
	W.N., Gott, J.R. III, Rhoads, J.E., Wang, Y., Bergeron, L.E., Golria,
	K.A., Long, D.C., Malhotra, S., \& Wanbsganss, J. 1997, \apj, 482, 75
\bibitem[Kundi\'c et al. (1997b)]{kun97b} Kundi\'c, T., Cohen, J.G., Blandford,
	R.D., \& Lubin, L.M. 1997, \aj, 114, 507
\bibitem[Lupie et al. 1997]{lup97} Lupie, O., Lallo, M., Cox, C., \& Bergreon,
	E. 1997, NICMOS Instrument Science Report 97-04, Space Telescope 
        Science	Institute
\bibitem[Madore et al. 1998]{mad98} Madore, B.F., et al. 1998, \nat, in press
\bibitem[Mao \& Schneider 1998]{mao98} Mao, S., \& Schneider, P. 1998, \aap, 
	in press
\bibitem[Maoz \& Rix 1993]{mao93} Maoz, D.. \& Rix, H.-W. 1993, \apj, 416, 425
\bibitem[McLeod 1998]{mck98} McLeod, B.A. 1998, in 1997 HST Calibration
	Workshop, ed S. Casertano et al., Space Telescope Science Institute,
	in press
\bibitem[McLeod \& Rieke]{mcl95} McLeod, K. K., \& Rieke, G. H. 1995, \apjl,
	454, L77
\bibitem[Poggianti 1997]{pog97} Poggianti, B.M. 1997, \aaps, 122, 399
\bibitem[Ramella, Pisani, \& Geller 1997]{ram97} Ramella, M., Pisani, A., \&
	Geller, M.J. \aj, 113, 483
\bibitem[Refsdal 1964]{ref64} Refsdal, S. 1964, \mnras, 128, 295
\bibitem[Rix et al. 1997]{rix97} Rix, H.-W., de Zeeuw, P.T., Cretton, N., van
	der Marel, R.P., \& Carollo, C.M. 1997, \apj, 488, 702
\bibitem[Romanowsky \& Kochanek 1998]{} Romanowsky, A. \& Kochanek, C. S. 
	1998, \apj\ in press (astro-ph/9708212)
\bibitem[Saha \& Williams 1997]{} Saha, P. \& Williams, L. L. R. 1997,
	\mnras, 292, 148
\bibitem[Schechter et al. (1997)]{sch97} Schechter, P.L., Bailyn, C.D., Barr, 
	R., et al. 1997, \apj, 475, L85
\bibitem[Schild \& Thomson 1997]{sci97} Schild, R.E., \& Thomson, D.J. 1997,
	\aj, 113, 130
\bibitem[Schlegel, Finkbeiner \& Davis 1998]{sfd98} Schlegel, D. J.,
        Finkbeiner, D. P. \& Davis, M. 1998, \apj\ in press 
        (astro-ph/9710327)
\bibitem[Schneider, Ehlers, \& Falco 1992]{sch92} Schneider, P., Ehlers, J.,
	and Falco, E.E. 1992, Gravitational Lenses (Springer-Verlag: Berlin)
\bibitem[Seljak 1994]{sel94} Seljak, U. 1994, \apj, 436, 509
\bibitem[Tonry 1998]{ton98} Tonry, J.L. 1998, \aj, 115, 1
\bibitem[Trotter 1998]{tro98} Trotter, C. 1998, Ph.D. Thesis, Massachusetts
	Institute of Technology
\bibitem[van der Marel 1991]{van91} van der Marel, R.P. 1991, \mnras, 253, 710
\bibitem[Vazdekis 1996]{vaz96} Vazdekis,~A., Casuso,~E., Peletier,~R.,
	Beckman,~J., 1996, \apjs, 106, 307.
\bibitem[Weymann et al. 1980]{wey80} Weymann, R.J., Latham, D., Angel, J.R.P.,
	Green, R.F., Liebert, J.W., Turnshek, D.A., Turnshek, D.E., \& Tyson,
	J.A. 1980, \nat, 285, 641
\bibitem[Young et al. (1981)]{you81} Young, P.J., Deverill, R.S., Gunn, J.E., 
	Westphal, J.A., \& Kristian, J. 1981, \apj, 244, 756
\bibitem[Zabludoff \& Mulchaey 1998]{zab97} Zabludoff, A.I., \& Mulchaey, 
	J.S. 1998, \apj, in press


\end{thebibliography}
\end{document}